\documentstyle[preprint,tighten,eqsecnum,floats,aps,epsfig]{revtex}

\begin{document}

\preprint{nucl-th/9608012}
\draft

\title{ Quasielastic $K^{+}$-nucleus scattering
}
\author{
A. De Pace 
}
\address{
 Istituto Nazionale di Fisica Nucleare, Sezione di Torino, \\
 via P.Giuria 1, I-10125 Torino, Italy
}
\author{
C. Garc\'\i a-Recio
}
\address{
Departamento de F\'\i sica Moderna, Faculdad de Ciencias, \\
Universidad de Granada, E-18071 Granada, Spain
}
\author{
E. Oset
}
\address{
Departamento de F\'\i sica Te\'orica and Instituto de F\'\i sica Corpuscular, \\
Centro Mixto Universidad de Valencia -- Consejo Superior de Investigaciones 
Cient\'\i ficas, \\
46100 Burjassot, Valencia, Spain
}
\date{August 1996}

\maketitle

\begin{abstract}
Quasielastic $K^+$-nucleus scattering data at $q=290$, $390$ and $480$ MeV/c are
analyzed in a finite nucleus continuum random phase approximation framework,
using a density-dependent particle-hole interaction. The reaction mechanism is
consistently treated according to Glauber theory, keeping up to two-step
inelastic processes. A good description of the data is achieved, also providing
a useful constraint on the strength of the effective particle-hole interaction
in the scalar-isoscalar channel at intermediate momentum transfers.
We find no evidence for the increase in the effective number of nucleons
participating in the reaction which has been reported in the literature.
\end{abstract}

\vfill

\eject

\section{ Introduction }
\label{sec:intro}

Quasielastic studies are traditionally a good source of information about
nuclear and nucleon structure. The main tool has been usually represented by
electron scattering experiments, since in this case the elementary probe-nucleon
reaction mechanism is regarded to be under better control. 
However, not all the possible nuclear response functions can be accessed in this
way and, furthermore, the extraction from the data of the response functions
that enter electron scattering can be model dependent.

Hadronic probes are an alternative source of quasielastic data. 
Among them it is particularly interesting the case of $K^+$-nucleus scattering,
since, for kaon laboratory momenta below 800 MeV/c, the elementary $K^+$-nucleon
($K^{+}N$) interaction is much weaker than other hadron-nucleon interactions. 
In fact, a weaker elementary interaction allows the projectile to penetrate
deeper inside the nucleus, thus probing regions of higher density and,
consequently, being more sensitive to collective phenomena.

Such an experiment has been performed at BNL, where quasielastic $K^+$-nucleus
double-differential cross sections have been measured for kaons with a
laboratory momentum of 705 MeV/c, using D, C, Ca and Pb as targets 
\cite{Kor93,Kor95a}. Data have been taken for scattering at 24$^\circ$, 
34$^\circ$ and 43$^\circ$, corresponding to approximately fixed momentum
transfers of 290, 390 and 480 MeV/c. In the case of C, the data have been taken
for all the momenta, whereas for Ca and Pb they are available at 290 and 480
MeV/c.

In Refs.~\cite{Kor93,Kor95a} the data have been analyzed using various
relativistic models to describe the nuclear dynamics; on the other hand, the
distortion of the incoming kaons has been accounted for employing a very simple
model, based on the effective number of nucleons participating in the reaction,
$N_{\text{eff}}$ (see Sec.~\ref{subsec:HP}).
The ``experimental'' values found for $N_{\text{eff}}$ are $\sim30\%$ higher
than the ones calculated in the Glauber theory. This finding is rather puzzling,
also in view of the fact that multiple scattering theory underestimates the
nuclear elastic scattering data \cite{Mar82,Sie84}, a fact which has been
interpreted as a signal of an enhancement of the in-medium $K^{+}N$ cross
section. As also noted in Ref.~\cite{Kor95a}, the quasielastic cross section is,
in general, proportional to the elementary single differential $K^{+}N$ cross
section and to the effective number of participating nucleons, the latter
depending, in turn, on the total $K^{+}N$ cross section (see 
Sec.~\ref{subsec:HP} for details). At the energy of the experiment, $K^{+}N$
scattering is nearly isotropic, so that the differential and total cross
sections are proportional; however, an increase in the $K^{+}N$ cross section
reduces $N_{\text{eff}}$ (and viceversa), leaving a quasielastic cross section
little dependent on the input $K^{+}N$ amplitudes.
Thus, the quasifree experiment on the one side does not shed light on the
elastic scattering data and, on the other, it seems to pose another puzzling
problem.

In this paper we would like to reanalyze the $K^+$-nucleus quasielastic data,
using a more realistic model for the reaction mechanism, --- already applied to 
($p,p'$) and ($p,n$) quasifree scattering \cite{DeP93,DeP95}, --- based on a
consistent implementation of Glauber theory for quasifree scattering.
The nuclear dynamics has been treated through a finite nucleus continuum random
phase approximation (RPA) calculation, also accounting for the coupling of the
particle-hole (ph) states to higher order configurations (spreading width).
The dynamics is non-relativistic, but the correct relativistic kinematics has
been used.

We will show that the puzzling experimental outcome for $N_{\text{eff}}$ is very
likely an artifact of the naive fitting procedure employed in the analysis of
the data and that a realistic (and {\em without free parameters}) theoretical
framework has no difficulties in giving a rather accurate description of the
experimental cross sections.

In Sec.~\ref{sec:TM} we introduce the theoretical methods employed in the
calculations: Here, one can find a brief description of the quasielastic nuclear
responses in finite nuclei; of the RPA using density-dependent interactions; of
the specific model for the ph effective potential that we have used; and of the
response functions to hadronic probes up to two-step processes in the Glauber
framework. In Sec.~\ref{sec:results}, we display the results of our calculations
and in the last Section we discuss their implications for the interpretation of
the data.

\section{ Theoretical methods }
\label{sec:TM}

\subsection{ Quasielastic nuclear response }
\label{subsec:QNR}

The nuclear response function to an external probe, transferring momentum $q$ 
and energy $\omega$, is proportional to the imaginary part of the polarization 
propagator\cite{Fet71}:
\begin{equation}
  R_\alpha(q,\omega) = -\frac{1}{\pi}\text{Im}
    \Pi_\alpha(\bbox{q},\bbox{q};\omega) \ .
\end{equation}
The latter reads
\begin{eqnarray}
  \Pi_\alpha(\bbox{q},\bbox{q}';\omega) = &&
    \sum_{n\ne0}\langle\psi_0|{\hat O}_\alpha(\bbox{q})|\psi_n\rangle
    \langle\psi_n|{\hat O}^\dagger_\alpha(\bbox{q}')|\psi_0\rangle
    \nonumber\\
  && \times\left[\frac{1}{\hbar\omega-(E_n-E_0)+i\eta}\right.
    \nonumber\\
  &&\quad\left.-\frac{1}{\hbar\omega+(E_n-E_0)-i\eta}\right] \ ,
\label{eq:Pi}
\end{eqnarray}
where \{$|\psi_n\rangle$\} is a complete set of nuclear eigenstates of energy
$E_n$, ${\hat O}_\alpha(\bbox{q})$ the second quantized expression of the
vertex operator and $\alpha$ labels the spin-isospin channel.
In the following, for the purpose of illustration, we shall deal explicitly only
with the scalar-isoscalar channel, --- which is, by the way, the dominant one in
$K^{+}N$ scattering at the energies of concern to us, --- where 
$O(\bbox{q},\bbox{r})=\exp(i\bbox{q}\cdot\bbox{r})$: The case of the spin modes,
which is slightly complicated by the spin algebra, is treated in detail in 
Ref.~\cite{DeP93}.

The angular part of $\Pi(\bbox{q},\bbox{q}';\omega)$ (we now drop, for
simplicity, the channel label $\alpha$) can be handled through a multipole
decomposition that reads
\begin{equation}
  \Pi(\bbox{q},\bbox{q}';\omega) = \sum_{JM}\Pi_J(q,q';\omega)
    Y^*_{JM}(\hat{\bbox{q}}) Y_{JM}(\hat{\bbox{q}}') \ ,
\end{equation}
so that
\begin{equation}
  R(q,\omega) = -\frac{1}{4\pi^2}\text{Im}\sum_{J}(2J+1)
    \Pi_J(\bbox{q},\bbox{q};\omega) \ .
\end{equation}
In a mean field (shell model) framework, one has for the particle-hole (ph)
polarization propagator:
\widetext
\begin{equation}
  \Pi_J^0(q,q';\omega) = \sum_{ph} Q^{JJ0}_{ph}(q)
    \left[\frac{1}{\omega-(\epsilon_p-\epsilon_h)+i\eta}
    -\frac{1}{\omega+(\epsilon_p-\epsilon_h)-i\eta}\right]
    {Q^{JJ0}_{ph}}^*(q') \ ,
\label{eq:PJ0}
\end{equation}
where $p$ ($h$) labels a complete set of single particle (hole) quantum numbers
and
\begin{equation}
  Q^{J\ell\sigma}_{ph}(q) =  \langle j_p j_h;J|\ell\sigma;J\rangle 
    (-i)^{\ell+1}(-1)^{\ell_h}2[4\pi(2\ell_p+1)(2\ell_h+1)]^{1/2} 
    {\cal I}_{\ell,ph}(q)\left(
      \begin{array}{ccc}
        \ell_p&\ell_h&\ell\\ 
        0&0&0
      \end{array}
    \right) \ .
\label{eq:QJl}
\end{equation}
\narrowtext
In (\ref{eq:QJl}), $\langle j_p j_h;J|\ell\sigma;J\rangle$ is the standard
$LS-jj$ recoupling coefficient and 
\begin{equation}
  {\cal I}_{\ell,ph}(q) = \int_0^\infty dr\,r^2j_\ell(qr)R_{p}(r)R_{h}(r) \ ,
\end{equation}
$R_{p(h)}(r)$ being the radial particle (hole) wave function and 
$\epsilon_{p(h)}$ the associated eigenvalues. They are obtained by solving the
Schroedinger equation with the Woods-Saxon potential
\begin{equation}
  W(r) = \frac{W_0}{1+e^{(r-R)/a}}
    +\left[\frac{\hbar c}{m_\pi^2 c^2}\right]^2\frac{W_{so}}{a r}
    \frac{e^{(r-R)/a}}{\left[1+e^{(r-R)/a}\right]^2}\,
    \bbox{\ell}\cdot\bbox{\sigma}
\end{equation}
(neglecting, for simplicity, the Coulomb term), where $m_\pi$ is the pion mass
and the following set of parameters has been employed:
\begin{equation}
  \begin{array}{rclrcl}
    W_0 &=& -54.8\,\text{MeV}, & \: W_{so} &=& -10\,\text{MeV} \ , \\
    R   &=& 1.27\,A^{1/2}\,\text{fm}, &\: a &=& 0.67\,\text{fm} \ .
  \end{array}
\end{equation}
Note that the sum appearing in Eq.~(\ref{eq:PJ0}) should be understood as a sum
over the discrete part of the spectrum and an integration over the continuum
one. Indeed, contrary to the widespread procedure of calculating the
polarization propagator in coordinate space and then Fourier transform to
momentum space, we have directly evaluated Eq.~(\ref{eq:PJ0}) in the latter.
Besides being fast and reliable, this procedure allows also for an important
extension to the shell model polarization propagator, namely the inclusion of
the {\em spreading width} of the ph states.

This can be accomplished by adding in Eq.~(\ref{eq:PJ0}) a complex ph 
self-energy, i.~e. through the following substitution:
\begin{equation}
  (\epsilon_p-\epsilon_h) \to (\epsilon_p-\epsilon_h)-\Sigma_{ph}(\omega) \ .
\end{equation}
Although $\Sigma_{ph}$ could in principle be calculated, we shall actually
employ a phenomenological parameterization, writing 
\begin{equation}
  \Sigma_{ph}(\omega)=
    \Delta_{ph}(\omega)+i\frac{\Gamma_{ph}(\omega)}{2} \ ,
\end{equation}
with
\begin{eqnarray}
  \Gamma_{ph}(\omega) &=&
    [\gamma_p(\hbar\omega+\epsilon_h)+\gamma_h(\epsilon_p-\hbar\omega)]
    (1+C_{TS}) \nonumber\\
  \\
  \Delta_{ph}(\omega) &=&
    [\Delta_p(\hbar\omega+\epsilon_h)+\Delta_h(\epsilon_p-\hbar\omega)]
    (1+C_{TS}) \ .
    \nonumber
\end{eqnarray}
The arguments of the functions $\gamma_{p(h)}$ and $\Delta_{p(h)}$ have been
inferred from the analysis of the second order particle and hole self-energy
contributions to $\Sigma_{ph}$.
The coefficients $C_{TS}$ represent the corrections, in each isospin $T$--spin
$S$ channel, due to the ph interference diagrams: These have been estimated 
\cite{Smi88} to contribute around 5\% in $S=T=1$ and $\approx20\div25\%$ in 
$S=0$, $T=1$ and $S=1$, $T=0$; given the smooth dependence of the quasielastic 
response on the ph self-energy and the small contribution of these modes to the
$K^+$-nucleus cross section, these corrections are not important and we shall 
set for simplicity $C_{TS}=0$ in these channels. Concerning the channel 
$S=T=0$, the estimate of Ref.~\cite{Smi88} gives $C_{00}=-1$, resulting in a 
complete cancellation of the ph spreading width. This outcome is valid in 
nuclear matter and in the limit of very large angular momenta, which is 
presumably good in the quasielastic peak (QEP) region.
However, at low transferred momenta the region of small energy transfers of the 
quasielastic response is dominated by resonances associated to specific (low)
angular momenta. In that case it is a better approximation to retain the full
width and, accordingly, we shall choose $C_{00}=-1+\exp[-(\omega/\omega_0)^2]$
with $\omega_0=30$ MeV: Then, resonances in the $10\div20$ MeV range get a width
of a few MeV, whereas for $\omega\gtrsim30$ MeV there is essentially no
spreading.

Finally, the $\gamma_{p(h)}$ are chosen according to the parametrization of 
Ref.~\cite{Smi88}, namely
\begin{eqnarray}
  \gamma_p(\epsilon) &=&
    2\alpha\left(\frac{\epsilon^2}{\epsilon^2+\epsilon^2_0}\right)
    \left(\frac{\epsilon^2_1}{\epsilon^2+\epsilon^2_1}\right)
    \theta(\epsilon)\nonumber\\
  \\
  \gamma_h(\epsilon) &=& \gamma_p(-\epsilon) \ , \nonumber
\end{eqnarray}
symmetrical with respect to the Fermi energy ($\epsilon_F=0$),
which gives a reasonable fit of the particle widths for
medium-heavy nuclei, using $\alpha=10.75$ MeV, $\epsilon_0=18$ MeV
and $\epsilon_1=110$ MeV \cite{Mah81}. The corresponding real parts are obtained
through a once-subtracted dispersion relation\cite{DeP93}.

A final remark, concerning relativistic kinematics effects on the response
functions, is in order. We shall be concerned with momentum transfers up to 500
MeV/c, where purely kinematical relativistic effects are starting to be sizable.
At $q=500$ MeV/c the relativistic position of the QEP is $\approx8$ MeV below
the non-relativistic one, the effect being even larger for energies on the right
hand side of the QEP. In Ref.~\cite{Alb93}, it had been shown that the
non-relativistic Fermi gas can be mapped into the relativistic one through the
simple prescription
\begin{equation}
  \omega \to \omega\left(1+\frac{\omega}{2m_N}\right) \ .
\label{eq:relativ}
\end{equation}
The validity of (\ref{eq:relativ}) in a finite nucleus shell-model calculation
has been checked in Ref.~\cite{Ama95}. The main effect of properly accounting
for the relativistic dispersion relation turns out to be a moderate shrink of
the response functions at the right of the QEP (see Sec.~\ref{sec:results}).

\subsection{ RPA with density-dependent interactions }
\label{subsec:RPA-DDI}

The polarization propagator introduced in the previous subsection is used in
this paper as input in a continuum RPA calculation of the response functions.

For a central density-independent interaction one has to solve, for each
multipole $J$, the following integral equation (see Ref.~\cite{Alb85} for the
complications introduced by the tensor interaction):
\begin{equation}
  \Pi_J^{\text{RPA}}(q,q';\omega) = \Pi_J^{0}(q,q';\omega) +
    \frac{1}{(2\pi)^3}\int_0^\infty dk\,k^2\Pi_J^{0}(q,k;\omega)
    V(k)\Pi_J^{\text{RPA}}(k,q';\omega) \ ,
\label{eq:RPA}
\end{equation}
$V(k)$ being the Fourier transform of the two-body potential.
However, realistic effective nuclear interactions are in general
density-dependent, especially in the scalar channels.

For a general interaction potential $V(\bbox{r},\bbox{r}')$, one can introduce
the double Fourier transform
\begin{eqnarray}
  V(\bbox{k},\bbox{k}') &=& \int d\bbox{r}\,d\bbox{r}'\, 
    e^{-i\bbox{k}\cdot\bbox{r}} e^{i\bbox{k}'\cdot\bbox{r}'}
    V(\bbox{r},\bbox{r}') \nonumber \\
  &=& \sum_J V_J(k,k')\,Y_{JM}(\hat{\bbox{k}}) Y^*_{JM}(\hat{\bbox{k}}') \ .
\label{eq:vddJ}
\end{eqnarray}
Then, instead of Eq.~(\ref{eq:RPA}), one has to solve the following RPA
equations:
\begin{eqnarray}
  \Pi_J^{\text{RPA}}(q,q';\omega) &=& \Pi_J^{0}(q,q';\omega) +
    \frac{1}{(2\pi)^6}\int_0^\infty dk\,k^2\int_0^\infty dk'\,{k'}^2 \,
    \Pi_J^{0}(q,k;\omega)V_K(k,k')\Pi_J^{\text{RPA}}(k',q';\omega) 
    \nonumber \\
  &=& \Pi_J^{0}(q,q';\omega) +
    \frac{1}{(2\pi)^3}\int_0^\infty dk'\,{k'}^2 
    K_J(q,k';\omega) \Pi_J^{\text{RPA}}(k',q';\omega) \ ,
\label{eq:RPA-dd}
\end{eqnarray}
having defined the kernel
\begin{equation}
  K_J(q,k';\omega) = \frac{1}{(2\pi)^3}\int_0^\infty dk\,k^2
    \Pi_J^{0}(q,k;\omega)V_J(k,k') \ .
\label{eq:kernel}
\end{equation}
Clearly, in the case $V(\bbox{r},\bbox{r}')\equiv V(\bbox{r}-\bbox{r}')$ one has
$V_J(k,k')=(2\pi)^3 V(k) \delta(k-k')/k^2$: The kernel (\ref{eq:kernel}) is then
reduced to $K_J(q,k';\omega)=\Pi_J^{0}(q,k';\omega)V(k')$ and one gets back to 
Eq.~(\ref{eq:RPA}).

From Eq.~(\ref{eq:RPA-dd}) it is apparent that the solution of the RPA equations
with density-dependent forces does not pose any additional technical problem,
apart from the input kernel, whose calculation is more involved.

As discussed in detail in the next subsection, in the calculations of 
Sec.~\ref{sec:results} we have been using a parameterization of the effective
nuclear interaction linear in the density. In general, in any (non-tensor)
channel one has
\begin{eqnarray}
  V(\bbox{r}_1,\bbox{r}_2) &=& V^{\text{ex}}(\bbox{r}_1-\bbox{r}_2) +
    V^{\rho}(\bbox{r}_1-\bbox{r}_2)\,\widetilde{\rho}
    \left(\frac{\bbox{r}_1+\bbox{r}_2}{2}\right) \nonumber \\
  &\equiv& V^{\text{ex}}(r) + V^{\rho}(r) \, \widetilde{\rho}(R) \ ,
\label{eq:vdd}
\end{eqnarray}
with $\bbox{r}=\bbox{r}_1-\bbox{r}_2$, $\bbox{R}=(\bbox{r}_1+\bbox{r}_2)/2$ and
$\widetilde{\rho}(R)=\rho(R)/\rho(0)$, $\rho(R)$ representing the nuclear
density, here approximated by the standard Fermi distribution.

In momentum space (\ref{eq:vdd}) reads 
\begin{equation}
  V(\bbox{k},\bbox{k}') = V^{\text{ex}}(k) (2\pi)^3 \delta(\bbox{k}-\bbox{k}')
    + V^{\rho}\left(\frac{\bbox{k}+\bbox{k}'}{2}\right) 
      \widetilde{\rho}(\bbox{k}-\bbox{k}')
\label{eq:vddq}
\end{equation}
and its $J$th component in the angular momentum expansion (\ref{eq:vddJ}) turns
out to be
\begin{equation}
  V_J(k,k') = V^{\text{ex}}(k) \frac{(2\pi)^3}{k^2} \delta(k-k') + 
    V^{\rho}_J(k,k')
\end{equation}
where
\begin{equation}
  V^{\rho}_J(k,k') = 2\pi\int d(\hat{\bbox{k}}\cdot\hat{\bbox{k}}') \,
    V^{\rho}\left(\frac{\bbox{k}+\bbox{k}'}{2}\right)
    \widetilde{\rho}(\bbox{k}-\bbox{k}') 
    P_J(\hat{\bbox{k}}\cdot\hat{\bbox{k}}') \ ,
\end{equation}
$P_J$ representing the ordinary Legendre polinomials.

\subsection{ Effective ph interaction }
\label{subsec:eff-int}

Two strategies are possible in order to determine the effective ph interaction
in the nuclear medium: One can either directly fix an effective potential by
fitting some phenomenological properties or start with a bare nucleon-nucleon
interaction and calculate the related $G$-matrix. 
Parameterizations of the ph interaction based upon the first procedure are
generally only available at very low momentum transfers (in terms of
Migdal-Landau parameters); since we are probing relatively high momenta, we have
resorted to use a $G$-matrix and we have chosen the calculation of 
Ref.~\cite{Nak84}, which, in our view, has the following appealing features:
It is based upon a realistic boson exchange potential; (nonlocal) exchange
contributions are included in the effective interaction, which is conveniently
parameterized in terms of Yukawa functions; a parameterization of the density
dependence is also provided.

The potential is given using the standard representation in spin and isospin
(no spin-orbit contribution will be considered in the following):
\begin{eqnarray}
  V(\bbox{k}_f,\bbox{k}_i;k_F) &=& F + 
     F' \bbox{\tau}_1\cdot\bbox{\tau}_2 +
     G  \bbox{\sigma}_1\cdot\bbox{\sigma}_2 +
     G' \bbox{\sigma}_1\cdot\bbox{\sigma}_2 \ \bbox{\tau}_1\cdot\bbox{\tau}_2
     \nonumber \\
  && +
     T  S_{12}(\hat{\bbox{q}}) +
     T' S_{12}(\hat{\bbox{q}}) \bbox{\tau}_1\cdot\bbox{\tau}_2 +
     H  S_{12}(\hat{\bbox{Q}}) +
     H' S_{12}(\hat{\bbox{Q}}) \bbox{\tau}_1\cdot\bbox{\tau}_2 \ ,
\end{eqnarray}
where $\bbox{q}=\bbox{k}_i-\bbox{k}_f$, $\bbox{Q}=\bbox{k}_i+\bbox{k}_f$ 
($\bbox{k}_i$, $\bbox{k}_f$ being the relative momenta in the initial and final 
state, respectively) and the coefficients are density and momentum dependent.

\begin{figure}[p]
\begin{center}
\mbox{\epsfig{file=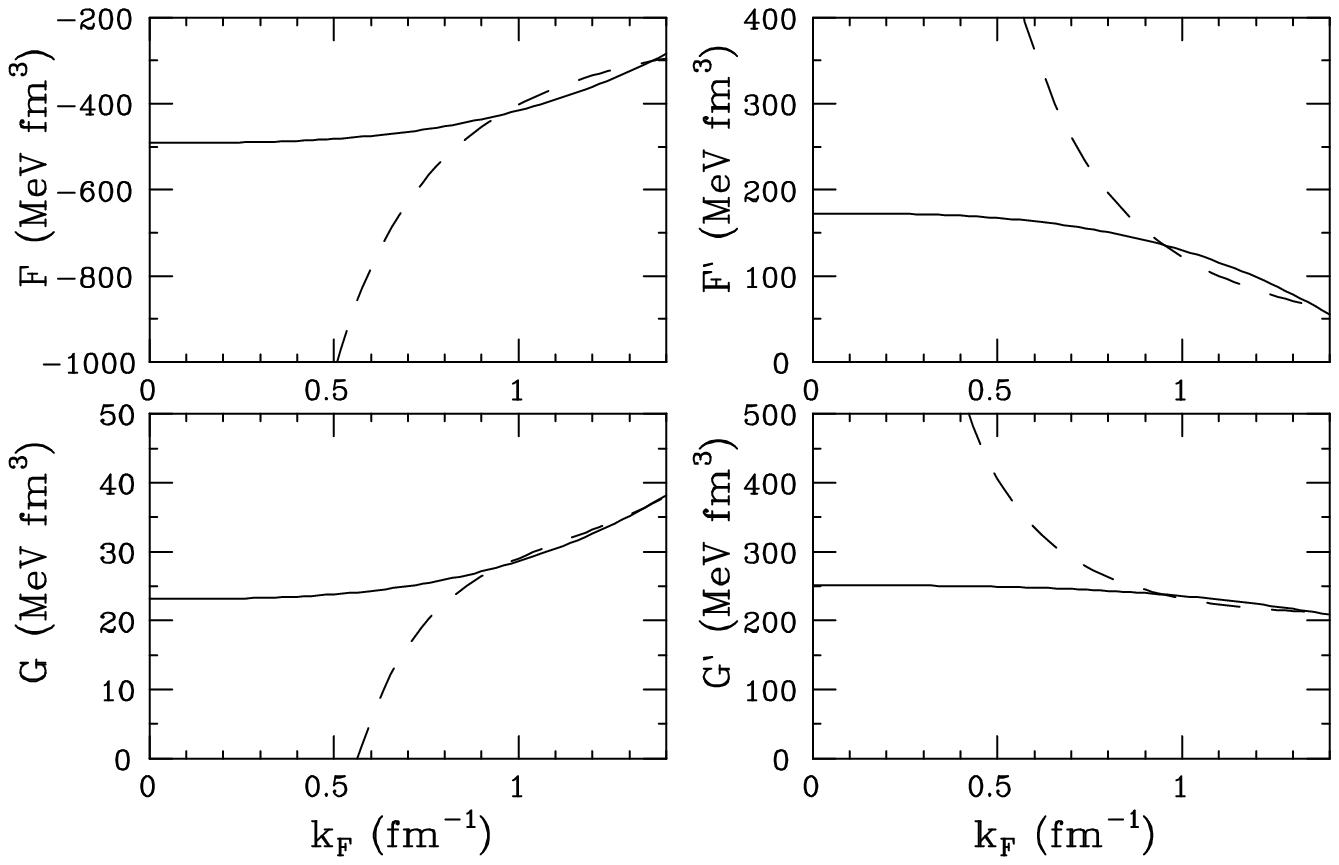,width=.85\textwidth}}
\caption{ Effective interaction in the non-tensor channels as a function of
$k_F$ at $q=0$; linear (solid) and from Ref.~\protect\cite{Nak84} (dashed)
density dependence.
  }
\label{fig:Veff-kF}
\vskip 3mm
\mbox{\epsfig{file=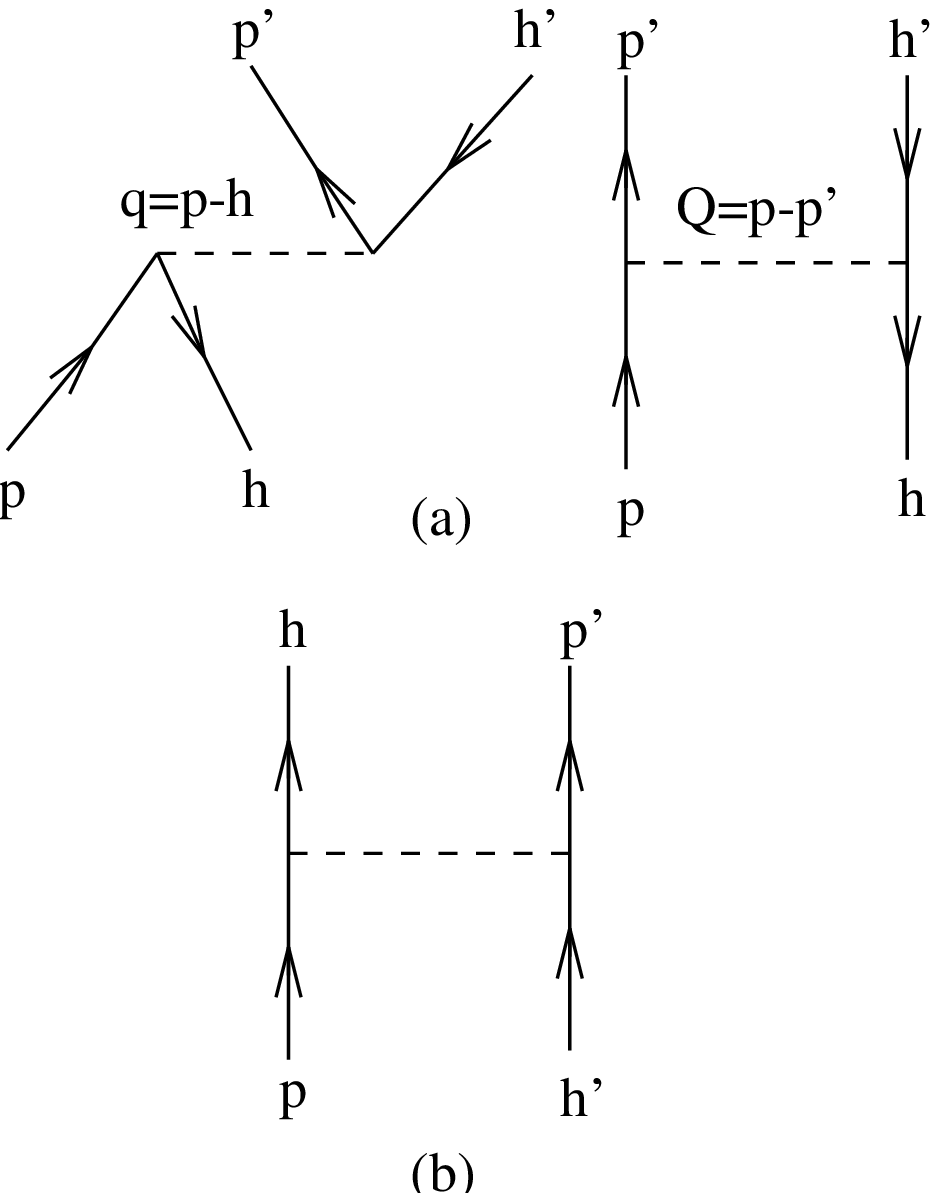,width=.50\textwidth}}
\caption{ (a) Direct and exchange ph matrix elements; (b) direct pp matrix 
element.
}
\label{fig:Veff-diag}
\end{center}
\end{figure}

However, before utilizing the nuclear matter interaction of Ref.~\cite{Nak84} in
a finite nucleus calculation of quasielastic responses, a few issues have to be
addressed.

a) The density dependence of the $G$-matrix is given in terms of density
dependent coupling constants, which is not very useful for our purposes.
Furthermore, the parameterization is fitted for $0.95 \text{fm}^{-1} < k_F <
1.36 \text{fm}^{-1}$, $k_F$ being the Fermi momentum, which spans a range of
densities down to roughly 1/3 of the central density: Extrapolation of their
parameterization to lower densities gives unreasonable results.
Thus, we have chosen to employ the linear $\rho$ dependence of 
Eq.~(\ref{eq:vdd}) [or (\ref{eq:vddq})], which is known to provide a reasonable
parameterization (see, e.~g., Ref.~\cite{Spe77}).
One can see in Fig.~\ref{fig:Veff-kF} a comparison of the two parameterizations
for the $k_F$ dependence of the effective interaction.
It should be noted that most of the contribution to the quasielastic responses
comes from densities where the two descriptions differ by a few per cent.

\begin{figure}[p]
\begin{center}
\mbox{\epsfig{file=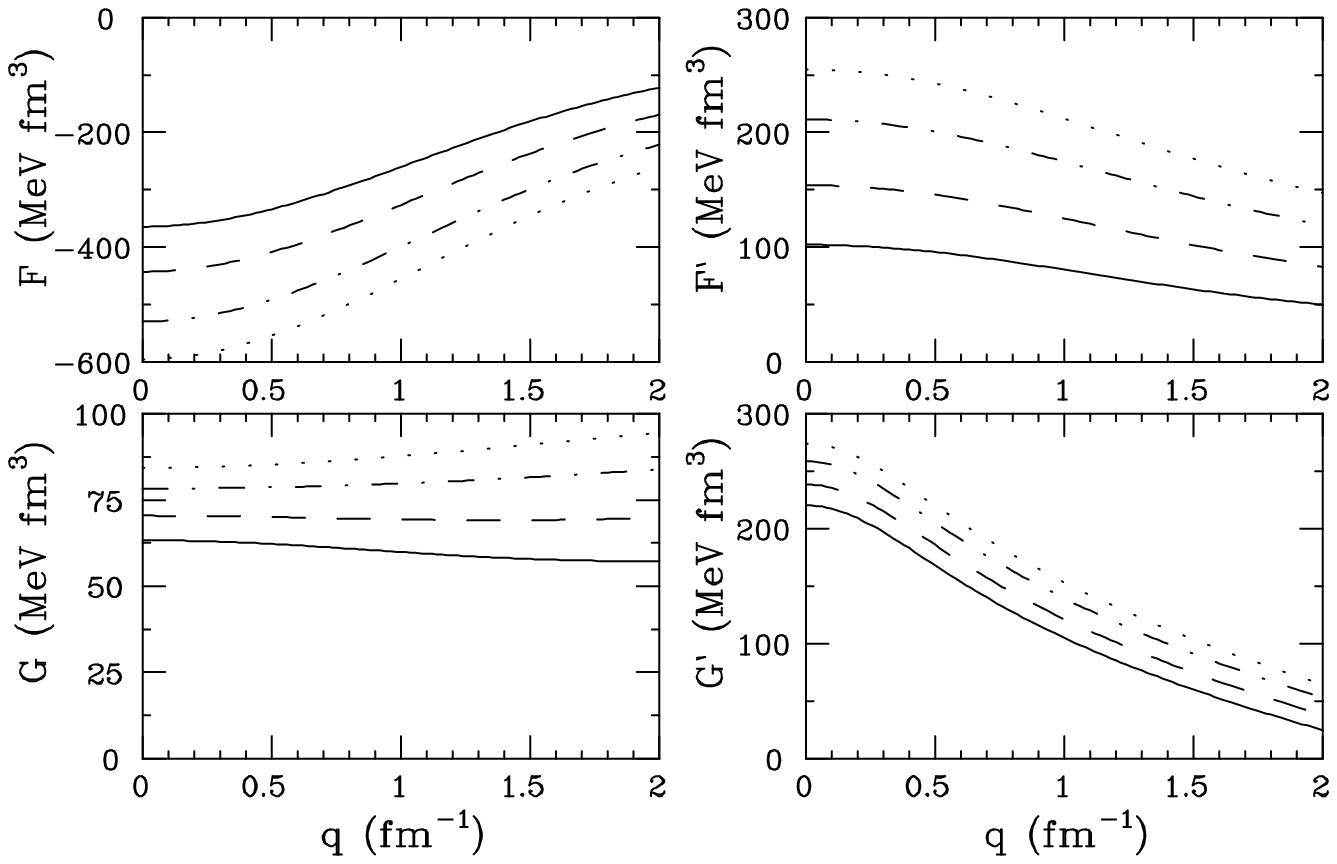,width=.85\textwidth}}
\caption{ Effective ph interaction in the non-tensor channels as a function of
$q$ at $k_F=1.36$ (solid), 1.25 (dashed), 1.10 (dot-dashed) and 0.95 fm$^{-1}$ 
(dotted).
}
\label{fig:Veff-nontensor}
\vskip 3mm
\mbox{\epsfig{file=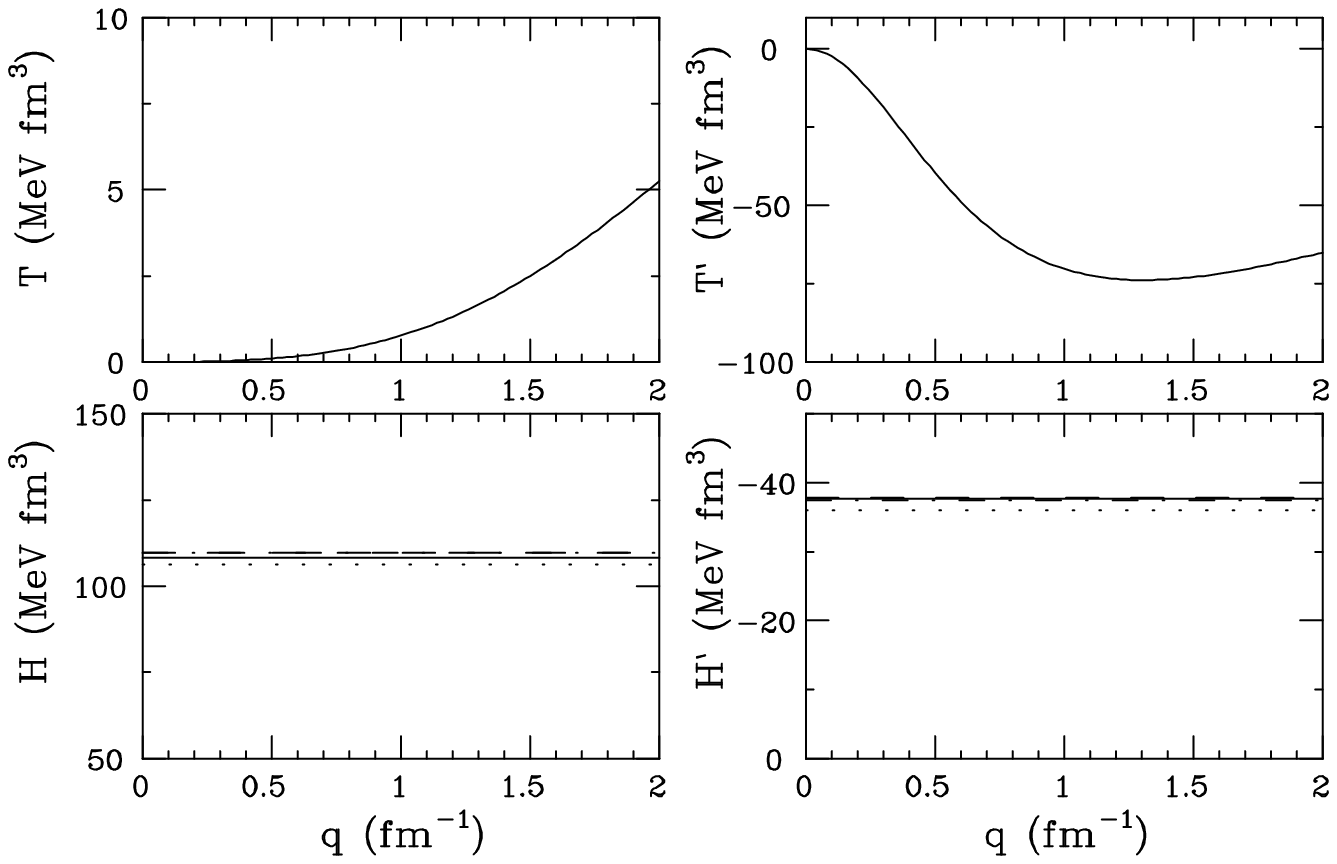,width=.85\textwidth}}
\caption{ As in Fig.~\protect\ref{fig:Veff-nontensor}, but for the tensor
channels; $T$ and $T'$ do not depend on the density.
}
\label{fig:Veff-tensor}
\end{center}
\end{figure}

b) In order to obtain, at a fixed density, a local interaction, one can use the
relation between $q$, $Q$ and $k_i$, i.~e., $Q=\sqrt{4k_i^2-q^2}$, substituting
$k_i$ with a suitably chosen average value, $\langle k_i \rangle$; then, the
only independent momentum is $q$.
The authors of Ref.~\cite{Nak84} were interested in a potential for nuclear
structure calculations: Hence, they put the momenta of the two nucleons in the
initial state on the Fermi surface and averaged over the relative angle, getting
$\langle k_i \rangle \approx 0.7 k_F$. Clearly, in this case one has the
constraint $0<q\lesssim1.4k_F$. 
On the other hand, we are interested in the ph interaction in the quasielastic
region: One nucleon in the initial state is below the Fermi sea, while the other
can be well above it. A look at Fig.~\ref{fig:Veff-diag} shows that $\bbox{k}_i$
is defined, in terms of the particle and hole momenta, as $\bbox{k}_i=
(\bbox{p}-\bbox{h}')/2=(\bbox{h}-\bbox{h}'+\bbox{q})/2$. Thus, at fixed 
$\bbox{q}$ one should average $k_i$ over $\bbox{h}$ and $\bbox{h}'$, getting 
$\langle k_i \rangle \approx \sqrt{6k_F^2/5+q^2}/2$.
Now $k_i$ is growing with $q$, so that there are no longer constraints on $q$,
and the exchange momentum turns out to be constant, $Q=\sqrt{6/5}k_F$.
One can see in Fig.~\ref{fig:Veff-nontensor} the resulting interaction in the
non tensor channels.

c) In the case of the tensor channels things are simpler, since in the
parameterization of Ref.~\cite{Nak84} there is no explicit density dependence
(Fig.~\ref{fig:Veff-tensor}). The coefficients of the exchange tensor operator,
$H$ and $H'$, display a very mild density dependence, induced by $Q$, which is
completely negligible. The only catch concerns the treatment of 
$S_{12}(\bbox{Q})$: Assuming that $\bbox{q}$ and $\bbox{Q}$ are orthogonal, with
some algebra one can show that 
$S_{12}(\hat{\bbox{Q}})=-S_{12}(\hat{\bbox{q}})/2$. 
Note, however, that $80\div90\%$ of the quasielastic cross section for $K^+$
scattering is due to the scalar-isoscalar channel.

\subsection{ Response functions with hadronic probes }
\label{subsec:HP}

The formalism introduced above is not enough when one is dealing with strongly
interacting probes, in which case a framework for the reaction mechanism must be
provided. For this purpose, we have chosen the Glauber approach\cite{Gla59},
including up to two-step inelastic processes.

Simple treatments within the Glauber theory usually amount to including the
effects due to rescattering in an effective number of nucleons participating in
the reaction, thus effectively renormalizing the response functions defined
above. In the case of $K^+$ quasielastic scattering, this approach has been
followed in Refs.~\cite{Kor93,Kor95a,Pie95}.

However, a consistent treatment within Glauber theory leads one to the
definition of {\em surface response functions}\cite{DeP93}.
A detailed derivation, in particular for the spin-isospin channel, can be found 
in Ref.~\cite{DeP93}. Here, we briefly sketch the relevant points, again using
the scalar-isoscalar channel as example.

A surface polarization propagator can be obtained from (\ref{eq:Pi}) by
substituting the vertex operator $O_{\alpha}(\bbox{q},\bbox{r})$, which 
describes the probe-nucleon coupling, with
\begin{eqnarray}
  O^{\text{surf}}_{\alpha}(\bbox{q},\bbox{r}) &=& \frac{1}{(2\pi)^2 
    f_{\alpha}(q)} \nonumber\\
  &&\times\int d\bbox{b}\,d \bbox{\lambda}\,
    e^{-\widetilde{\sigma}_{\text{tot}} T(b)/2}
    e^{i(\bbox{q}-\bbox{\lambda})\cdot\bbox{b}}f_{\alpha}(\lambda)
    \nonumber\\
  &&\times O_{\alpha}(\bbox{\lambda},\bbox{r}) \ ,
\end{eqnarray}
$\bbox{b}$ and $\bbox{\lambda}$ being bidimensional vectors in the plane
orthogonal to the direction of motion of the projectile, $f_{\alpha}$ the
elementary probe-nucleon amplitudes, $\widetilde{\sigma}_{\text{tot}}$ the
effective total probe-nucleon cross section and 
\begin{equation}
  T(b) = \int_{-\infty}^{+\infty} dz\,\rho(r=\sqrt{b^2+z^2}) \ ,
\end{equation}
$\rho(r)$ being the nuclear density. With these definitions, the effective
number of participating nucleons is given as 
\begin{equation}
  N_{\text{eff}} = \int d\bbox{b}\, T(b)e^{-\widetilde{\sigma}_{\text{tot}}
    T(b)} \ .
\end{equation}
$\widetilde{\sigma}_{\text{tot}}$ depends on the energy and we shall take the 
average over the total cross sections at the incoming and outgoing projectile 
energies: For $K^+$ scattering at $k=705$ MeV/c, the variation of 
$\widetilde{\sigma}_{\text{tot}}$ is rather weak, going roughly from 14 mb to 13
mb for energy losses up to 250 MeV.

One then finds, for the $J$-th multipole,
\widetext
\begin{eqnarray}
  \Pi^{\text{surf}}_{J(\alpha)}(q,q;\omega) &=& \Pi_{J(\alpha)}(q,q;\omega)
  \nonumber\\
  && +\frac{1}{|f_{\alpha}(q)|^2}\int_0^\infty
  d\lambda\,\lambda\int_0^\infty
  d\lambda'\,\lambda'\,\text{Re}[f^*_{\alpha}(\lambda)\,
  f^{\phantom{*}}_{\alpha}(\lambda')\,
  G^{(0)}_J(\lambda,\lambda';q)] \Pi_{J(\alpha)}(\lambda,\lambda';\omega)
  \nonumber\\
  && -2\frac{1}{|f_{\alpha}(q)|^2}\int_0^\infty
  d\lambda\,\lambda\,\text{Re}[f^*_{\alpha}(q)\,
  f^{\phantom{*}}_{\alpha}(\lambda)\,
  H^{(0)}_J(\lambda;q)]\Pi_{J(\alpha)}(q,\lambda;\omega) \ ,
\end{eqnarray}
\narrowtext
having set 
\begin{mathletters}
\begin{eqnarray}
  G^{(0)}_J(\lambda,\lambda';q) &=& \sum_{m} c_{Jm}
    g^*_m(\lambda,q) g_m(\lambda',q) \\
  H^{(0)}_J(\lambda;q) &=& \sum_{m} c_{Jm}g_m(\lambda,q) \ ,
\end{eqnarray}
\end{mathletters}
where
\begin{equation}
  g_m(\lambda,q)=\int_0^\infty  db\,b 
    \left[1-e^{-\widetilde{\sigma}_{\text{tot}} T(b)/2}\right]
    J_m(\lambda b)\,J_m(qb)
\end{equation}
and
\begin{eqnarray}
  c_{Jm} &=& I_{J+m} \frac{(J-m-1)!!(J+m-1)!!}{(J+m)!!(J-m)!!} \nonumber\\
  \\
  I_{J+m} &=& \left\{
    \begin{array}{ccc}
      0,& J+m &\text{odd} \\ 
      1,& J+m &\text{even}
    \end{array}
    \right. \ .
  \nonumber
\end{eqnarray}
With the previous definitions, the one-step surface response functions are given
as $R_{\alpha}^{\text{surf}}(q,\omega)=-\text{Im}\sum_J(2J+1)
\Pi_J^{\text{surf}}(q,q;\omega)/4\pi^2$ and the double differential cross
section turns out to be
\begin{equation}
  \left.\frac{d^2\sigma}{d\Omega d\epsilon'}\right|_{\text{1-step}} =
    \sum_{\alpha}|f_{\alpha}(q)|^2 R_{\alpha}^{\text{surf}}(q,\omega) \ ,
\end{equation}
to be compared with the effective number approximation, which reads
\begin{equation}
  \left.\frac{d^2\sigma}{d\Omega d\epsilon'}\right|_{\text{1-step}} =
    N_{\text{eff}} \sum_{\alpha}|f_{\alpha}(q)|^2 R_{\alpha}(q,\omega) \ .
\label{eq:Neff-approx}
\end{equation}
Multiple scattering inelastic contributions can, in principle, be incorporated
along the same lines. Since in the kinematic regions of interest they turn out
to be much smaller than the one-step terms, it is not worth going through a very
complex formalism and one can safely stick to the effective number
approximation. In this case, the two-step contribution is proportional to the
convolution of two one-step response functions\cite{Smi87}.
By summing over all the possible channels and using free one-step response
functions, one can define the two-step response as
\begin{eqnarray}
  R^{(2)}(q,\omega) = &&
    \frac{{\cal D}_2}{k^2}\frac{1}{\sum_{ST}|f_{ST}(q)|^2}
    \int d\bbox{q}' \int_0^\omega d\omega' \nonumber\\
  &&\left\{
    \sum_{ST}|f_{ST}(q')|^2 R^{(0)}(q',\omega')
    \sum_{ST}|f_{ST}(|\bbox{q}-\bbox{q}'|)|^2
    R^{(0)}(|\bbox{q}-\bbox{q}'|,\omega-\omega') \right.\nonumber \\
  && \left. \ +2
    \sum_{S}|f_{S1}(q')|^2 R^{(0)}(q',\omega')
    \sum_{S}|f_{S1}(|\bbox{q}-\bbox{q}'|)|^2
    R^{(0)}(|\bbox{q}-\bbox{q}'|,\omega-\omega') \right\}\ , \nonumber \\
\label{eq:R2}
\end{eqnarray}
where $k$ is the momentum of the projectile, $f_{ST}$ the $KN$ amplitude in the
$S,T$ channel and
\begin{equation}
  {\cal D}_2=\frac{1}{2}\int  d\bbox{b}\,T^2(b)
    e^{-\widetilde{\sigma}_{\text{tot}}T(b)}
\label{eq:D2}
\end{equation}
is proportional to the effective number of pairs participating in the double
scattering.

\begin{figure}[t]
\begin{center}
\mbox{\epsfig{file=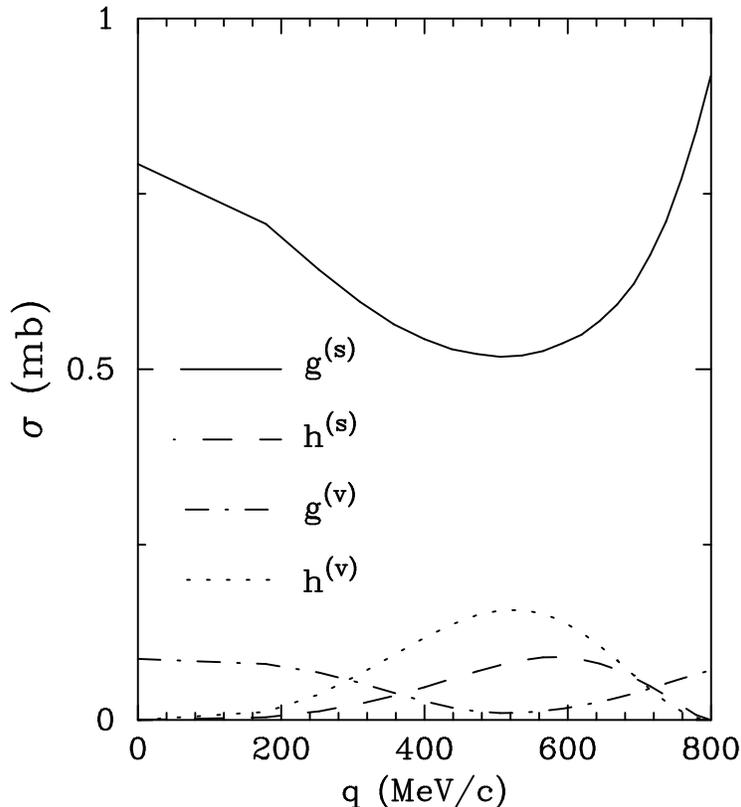}}
\caption{ Center-of-mass $K^+ N$ squared amplitudes corresponding to a
$K^+$ laboratory momentum $k=705$ MeV/c.
  }
\label{fig:amplitudes}
\end{center}
\end{figure}

The full quasielastic $K^+$-nucleus cross section discussed in 
Sec.~\ref{sec:results} is given by the sum of one- and two-step terms:
\begin{equation}
  \frac{d^2\sigma}{d\Omega d\epsilon'} =
    \sum_{\alpha}|f_{\alpha}(q)|^2 \left[R_{\alpha}^{\text{surf}}(q,\omega) +
    R^{(2)}(q,\omega) \right] \ .
\end{equation}

We conclude this Section with a few remarks about the elementary $K^+ N$
amplitudes employed in the calculations.
The amplitude for elastic $K^+ N$ scattering can be written as 
$f^{(s)}+f^{(v)}\tau_3$, with 
$f^{(i)}=g^{(i)}+i\bbox{\sigma}\cdot\hat{\bbox{n}}\,h^{(i)}$, 
$\hat{\bbox{n}}$ being a unit vector normal to the scattering plane.
In Fig.~\ref{fig:amplitudes} one can see the amplitudes as a function of the
momentum transfer for $k=705$ MeV/c\cite{Arn78,Mar75}: The dominant channel is 
clearly the scalar-isoscalar one.

Some care should be taken in the choice of the reference frame where the $K^+ N$
amplitudes are evaluated. In order to be able to factorize the two-body
amplitude out of the nuclear response functions, an ``optimal'' choice of the
reference frame has to be done\cite{Gur86,Smi89}: The optimal momentum of the
struck nucleon turns out to be
\begin{equation}
  \bbox{p}_{\text{opt}} = -\frac{\bbox{q}}{2}\left(1-
    \frac{\omega}{q}\sqrt{1+\frac{4m_N^2}{q^2-\omega^2}}\right) \ ,
\end{equation}
such as to reduce to zero at the QEP and to the Breit frame value at $\omega=0$.

The most relevant consequence of this choice of frame is that the elementary
projectile-nucleon amplitudes should be evaluated at an effective laboratory
kinetic energy defined by 
\begin{equation}
  T_L^{\text{eff}} = \frac{E_k E_{\text{opt}}-\bbox{k}\cdot\bbox{p}_{\text{opt}}
    -m_p m_N}{m_N} \ ,
\end{equation}
$m_p$ and $k$ being the projectile mass and momentum, respectively.
This introduces a dependence on $\omega$ of the amplitudes entering in the
evaluation of the quasielastic cross section, which can be important  if the
former are strongly energy dependent: At the kinematics of relevance to us, the
effective $K^+$ momentum varies roughly in the range $500\div800$ MeV/c, where
the amplitudes change sufficiently slowly to make the effects not at all
dramatic.

\section{ Results }
\label{sec:results}

\begin{figure}[tb]
\begin{center}
\mbox{\epsfig{file=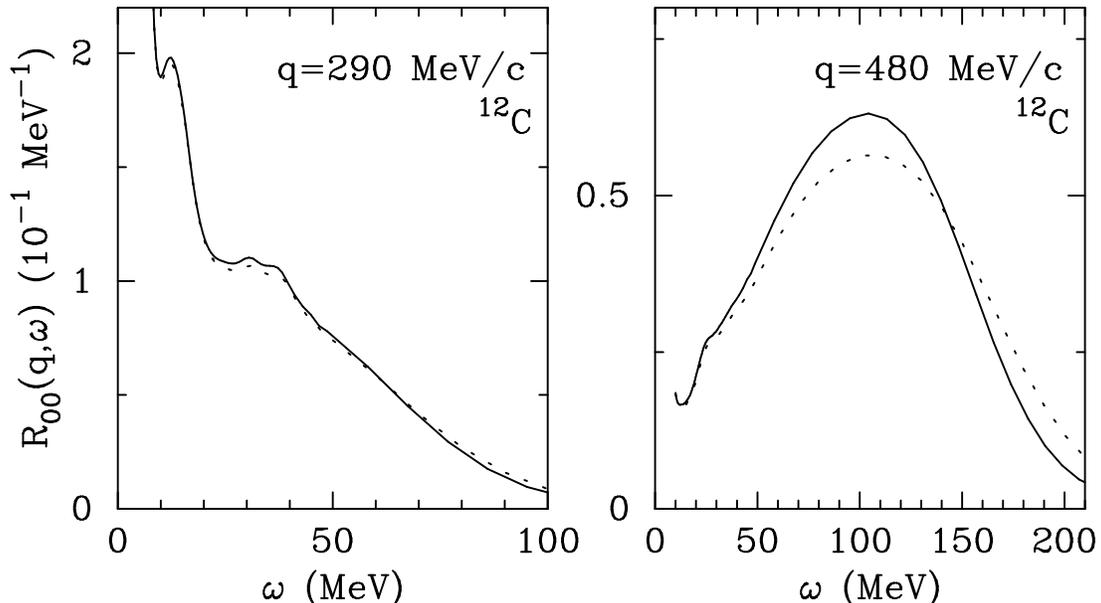}}
\caption{ Scalar-isoscalar RPA response function for $K^+$ quasielastic 
scattering: Non-relativistic (dotted) and with the kinematical relativistic
corrections discussed in the text (solid).
  }
\label{fig:rel}
\end{center}
\end{figure}

Let us start by briefly discussing the relativistic kinematical effects,
introduced at the end of Sec.~\ref{subsec:QNR}.
In Fig.~\ref{fig:rel} one can see the RPA scalar-isoscalar response of $^{12}$C
to $K^+$ probes in the non-relativistic case compared to the response function
where the relativistic dispersion relation has been accounted for.
As expected, the effect is negligible at the lowest momentum transfer ($q=290$
MeV/c), whereas at higher momenta ($q=480$ MeV/c) it produces a moderate shrink
of the response at the right of the QEP, compensated by an enhancement
($\sim10\%$) at the peak position.

Next, we would like to discuss the uncertainties connected to the choice of the
effective ph interaction for quasielastic calculations. 
The knowledge of the latter suffers, in general, from many shortcomings: When it
is theoretically calculated, --- as the one we employ here (see 
Sec.~\ref{subsec:eff-int}), --- sources of uncertainties arise because of the
specific many-body scheme which has been adopted, from relativistic effects (see
Ref.~\cite{Nak87} for an effective interaction along the lines of 
Ref.~\cite{Nak84}, but in a relativistic context) and from the approximations 
made in the actual calculations or in extrapolating to the ph kinematical regime
(see Sec.~\ref{subsec:eff-int}); when the effective interaction is just fitted 
to phenomenological properties, it can suffer from ambiguities in the
parameterization (many sets of parameters reproducing the same body of data) and
from the limited range of momenta which is covered (see, for instance, 
Ref.~\cite{Spe77} for a phenomenological density-dependent interaction expressed
in terms of Migdal-Landau parameters at $q\approx0$).

On the other hand, one can of course reverse the argument and use the
quasielastic studies to gain insight into the effective ph interaction at
momentum transfers of a few hundreds MeV. One obvious difficulty in this case is
related to the fact that it is not always possible to disentangle the various
spin-isospin channels: The only reaction for which this has been achieved is the
($\vec{p},\vec{n}$) charge-exchange one, where the separated isovector
spin-longitudinal and spin-transverse responses have been extracted
\cite{Tad94}. However, the strong interaction of protons with nuclei constrains 
that reaction to the low density peripheral region of the nucleus, making it 
little sensitive to RPA effects; moreover, while the spin-longitudinal
channel can be well described, the transverse one shows much more strength than
expected \cite{DeP95}. In this regard, the $K^+$-nucleus reaction is much more
promising for two reasons: First of all, the small $K^+ N$ cross section makes
the kaon enter inside the nucleus much more deeply than protons or pions, in
turn implying stronger collective effects; secondly, as seen in 
Fig.~\ref{fig:amplitudes}, the scalar-isoscalar channel is largely dominant,
making this reaction a ``quasipure'' probe of the $S=T=0$ mode.

\begin{figure}[tb]
\begin{center}
\mbox{\epsfig{file=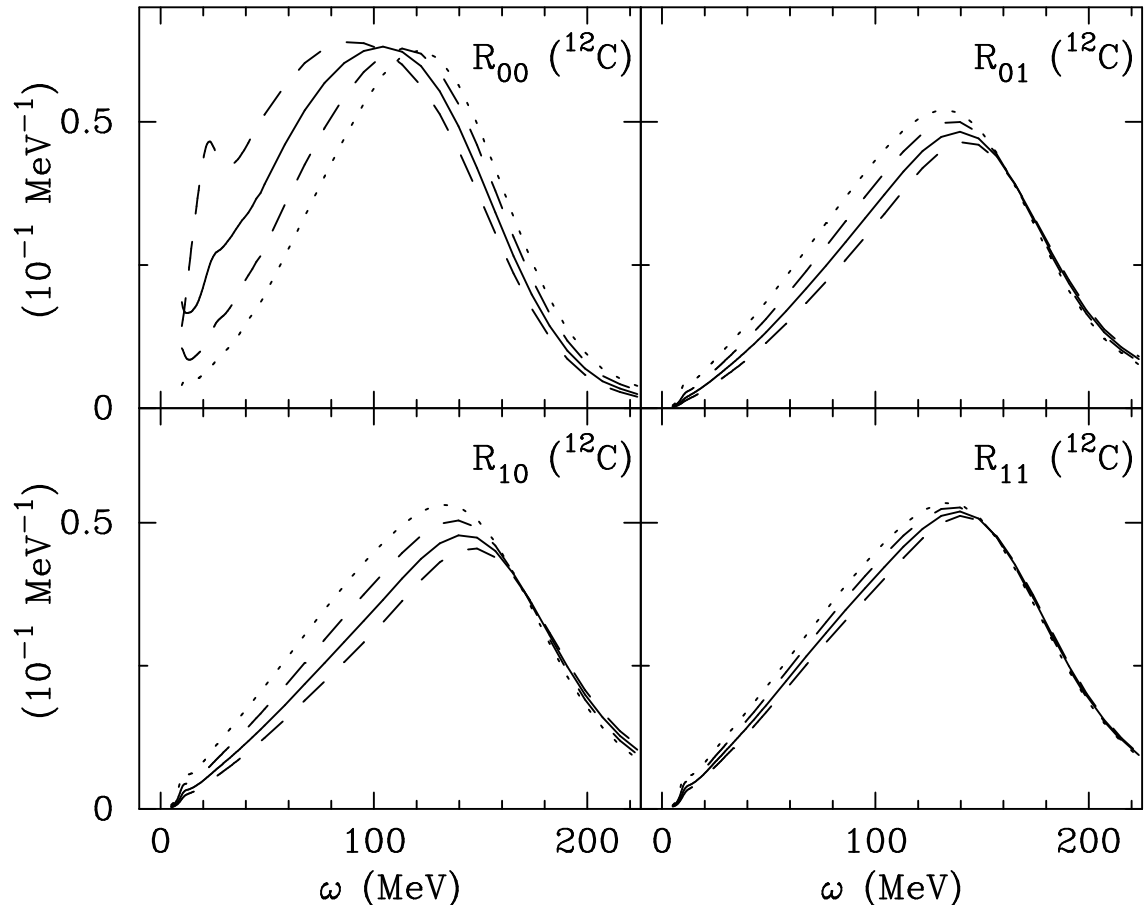}}
\caption{ $K^+$-$^{12}$C quasielastic response functions $R_{ST}$ ($S,T=0,1$) at
$q=480$ MeV/c. The dotted line represents the uncorrelated case; the solid line
the RPA responses corresponding to the interaction of 
Sec.~\protect\ref{subsec:eff-int}; the dashed lines the RPA responses with the
same interaction scaled by $\pm50\%$.
  }
\label{fig:RPA}
\end{center}
\end{figure}

Whatever be the attitude towards this problem, we believe it is useful to gain
some feeling about the sensitivity of the collective effects to the input
effective potential. For this purpose, we display in Fig.~\ref{fig:RPA}, for the
four spin-isospin channels, the surface RPA response functions at $q=480$ MeV/c,
calculated with the ph interaction discussed in Sec.~\ref{subsec:eff-int},
comparing them to the RPA responses calculated with the same interaction scaled
by $\pm50\%$. This amount for the rescaling is not an estimate of the
theoretical or phenomenological uncertainty of the ph interaction; however,
concerning the scalar-isoscalar channel, there are reasons to believe that the
$G$-matrix estimate of Ref.~\cite{Nak84} gives too much attraction: As discussed
by the authors, the extracted $f_0$ Landau parameter makes nuclear matter
unstable, a shortcoming that can be cured by higher order 
contributions\footnote{ Also relativistic effects give rise to a weaker
attraction: In Ref.~\protect\cite{Nak87} the strength in the $S=T=0$ channel has
been shown to be reduced by nearly a factor $1/2$ at $q=0$; however, in the
range of momenta of interest to us, the relativistic and non-relativistic
$G$-matrices appear to be comparable.}.
Thus, while the case with the interaction increased by $50\%$ is taken mainly 
for illustrative purposes, the case with a weaker interaction is somewhat more
realistic.

By inspecting Fig.~\ref{fig:RPA} one can note the following:

i) the dependence of the responses on the potential is somehow desensitized by
the RPA chain: For the $S=T=0$ mode the variation of the response function at
the left of the QEP is typically about $20\div30\%$, apart from the very low
energy region ($\omega\lesssim30$ MeV), where it grows up to $\sim50\%$; for the
other modes, still at the left of the QEP, the variation is always much smaller,
say $\lesssim5\div10\%$;

ii) at the right of the QEP the effect of correlations is much smaller, being
limited to a few per cent for the $S=T=0$ mode and to practically nothing for 
the other modes;

iii) the relatively weak sensitivity of the isovector and spin channels to 
variations of the input potential makes the small ($10\div20\%$) contamination 
from these modes in the $K^+$-nucleus cross section rather stable with respect 
to uncertainties due to correlations, enhancing the argument in favour of this
reaction as a probe of the scalar-isoscalar channel.

The results if Fig.~\ref{fig:RPA} are given for $q=480$ MeV/c: Similar
considerations apply also to the lowest momenta.

\begin{figure}[p]
\begin{center}
\mbox{\epsfig{file=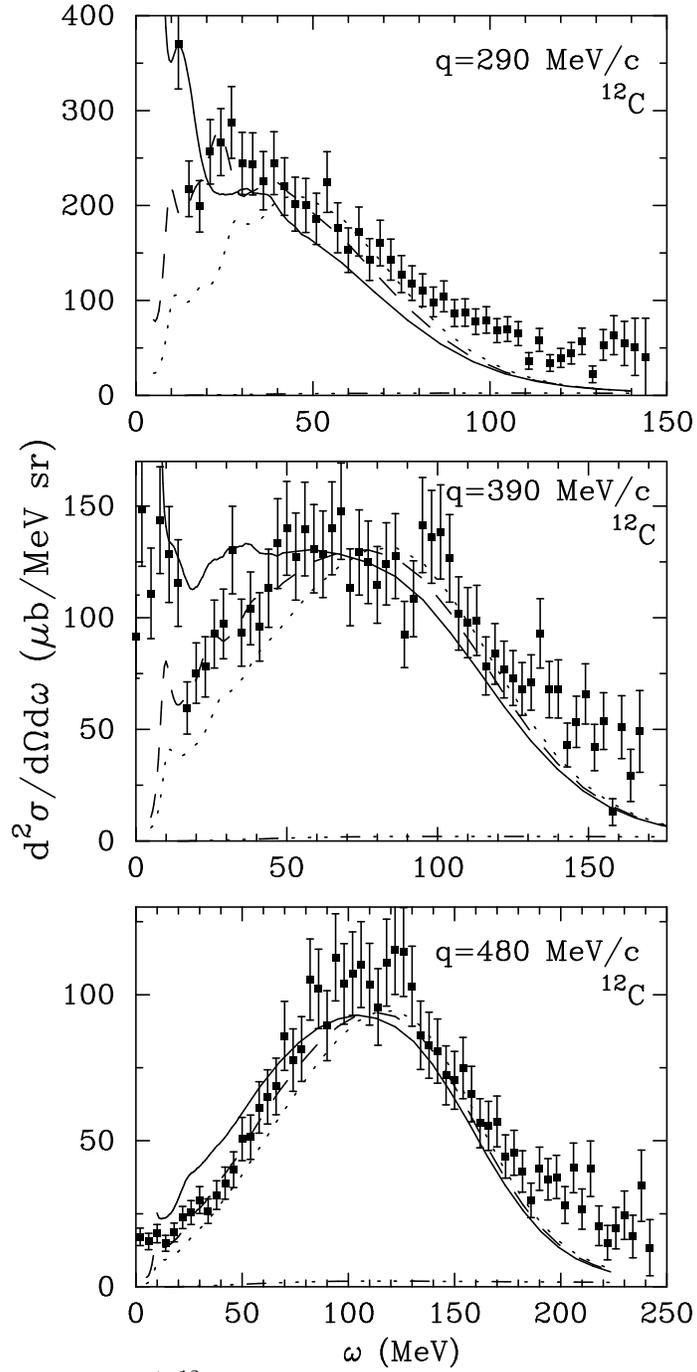,height=.8\textheight}}
\caption{ Cross section for $K^+$-$^{12}$C quasielastic scattering: Free 
response (dotted), RPA with the interaction of 
Sec.~\protect\ref{subsec:eff-int} (solid); RPA with the same interaction reduced
by $50\%$ (dashed). In all the cases, the two-step contribution has been added 
and it is also shown separately (dot-dashed).
  }
\label{fig:C12}
\end{center}
\end{figure}

In Fig.~\ref{fig:C12} one can see the data for the quasielastic $K^+$-C cross
section \cite{Kor93,Kor95a} compared to our calculations. In the theoretical
cross sections, besides the case with free response functions (dotted), there 
are included the nuclear responses corresponding to the interaction of 
Sec.~\ref{subsec:eff-int} (solid) and to a reduction of $50\%$ of the same
interaction (dashed); the curves include the two-step contribution, which is 
also shown separately (dot-dashed).

From inspection of the figure, one sees that the strength of the cross sections
on the right hand side of the QEP is well reproduced by both the correlated and
the free responses (apart from the very high energy tail) at all the momentum
transfers; however, at the lowest momenta the data show a clear distortion of
the typical quasielastic shape, which is not compatible with the uncorrelated
cross sections. 

The experimental energy resolution is not sufficient to cut out the elastic
scattering contribution; moreover, low-lying discrete nuclear excitations are
much affected by the detail of the nuclear model (such as shell model parameters
and spreading width). If, for these reasons, one excludes the very low energy
tail (say, $\omega\lesssim15$ MeV) it appears that the model with a weaker ph
interaction gives a remarkably good description of the data at $q=290$ and $390$
MeV/c.

At the highest momentum, the elastic contamination and the discrete excitations
have been washed out; the interaction in both the correlated models is
sufficiently weak to cause little distortion of the quasifree shape, but the low
energy part of the spectrum seems again to favour the model with the weaker ph
interaction.

It is worth noticing the smallness of the two-step contribution, which is at 
most a few per cent of the one-step term at $q=480$ MeV/c: This can be 
contrasted, for instance, to the isovector spin-transverse response to 500 MeV 
protons, where one finds, at $q\cong500$ MeV/c, a sizeable $20\div30\%$ 
contribution from two-step processes \cite{DeP95}. Protons of 500 MeV of
kinetic energy have a squared momentum which is roughly twice
the one of 700 MeV/c kaons; furthermore, the two-step factor ${\cal D}_2$ of
Eq.~(\ref{eq:R2}) and (\ref{eq:D2}) is larger by roughly a factor two for the 
kaon reaction, --- because of the smaller probe-nucleon total cross section
$\widetilde{\sigma}_{\text{tot}}$, --- but this is compensated by the factor two
arising from the two possible orderings of the charge-exchange reaction (see
formula (8) of Ref.~\cite{DeP95}). Hence, apart from a factor 2 in the
coefficient, the relative weight of the two-step
term in the two reactions has to be driven by the elementary amplitudes
entering (\ref{eq:R2}) (the response functions being in all the cases the free
ones), that is by the factor $|f_{00}(|\bbox{q}-\bbox{q}'|)|^2 |f_{00}(q')|^2 / 
|f_{00}(q)|^2$ in (\ref{eq:R2}) (since the scalar-isoscalar channel is dominant)
and by the factor
$|f^{NN}_{00}(|\bbox{q}-\bbox{q}'|)|^2 |f^{NN}_{T}(q')|^2 / |f^{NN}_{T}(q)|^2$ 
in the analogous expression for the ($p,n$) reaction, $\bbox{q}$ being the
external momentum and $\bbox{q}'$ the integration variable, whereas $f^{NN}_{T}$
and $f^{NN}_{00}$ are the isovector spin-transverse and the scalar-isoscalar
$NN$ amplitudes, respectively.

\begin{figure}[tb]
\begin{center}
\mbox{\epsfig{file=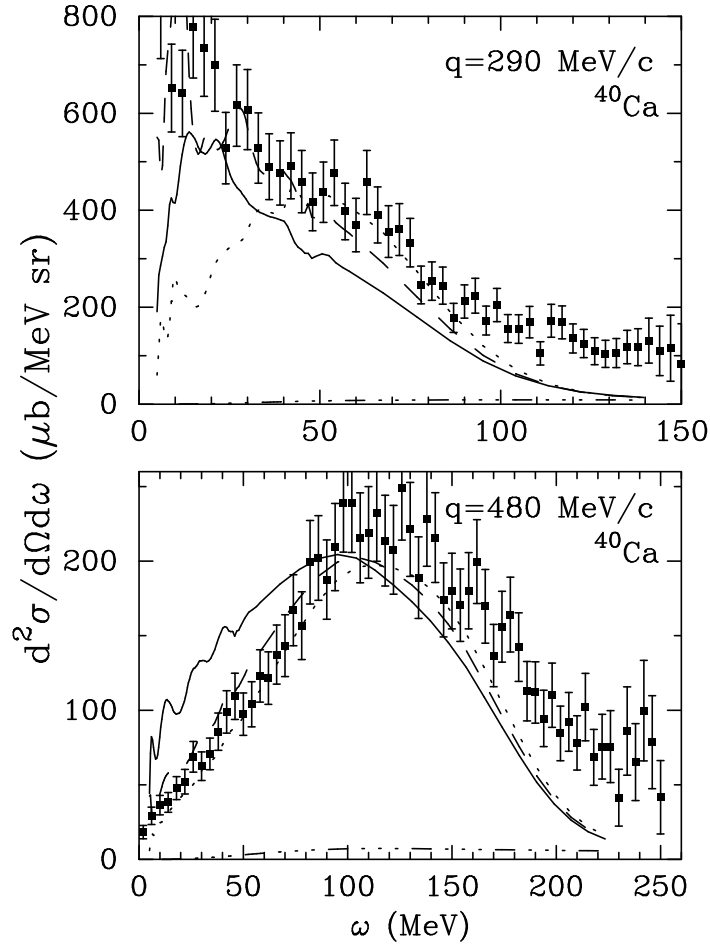,height=.55\textheight}}
\caption{ As in Fig.~\protect\ref{fig:C12}, but for $^{40}$Ca.
  }
\label{fig:Ca40}
\end{center}
\end{figure}

The different size of the two-step contributions is indeed determined by the 
different strength and the different momentum behaviour of the above amplitudes:
In the kaon case, the scalar-isoscalar amplitude has little momentum dependence,
so that $|f_{00}(q')|^2 / |f_{00}(q)|^2$ is roughly between one and two, while
the remaining squared amplitude is on average $\sim1.5$~mb (in the laboratory);
on the other hand, for the ($p,n$) reaction
at $q\cong500$ MeV/c, the ratio $|f^{NN}_{T}(q')|^2 / |f^{NN}_{T}(q)|^2$ is
typically in the range $4\div8$, whereas the remaining amplitude gives
contributions mainly in the range from 10 to 22 mb.
In qualitative words, we can say that because the $NN$ cross section is
relatively forward peaked, --- unlike the $K^+ N$ one, --- a double scattering
in which the momentum transfer $q$ is shared by the two nucleons is favored in
the $NN$ case with respect to the $K^+ N$ one.

Finally, we display also the available data for Ca at $q=290$ and $480$ MeV/c
(Fig.~\ref{fig:Ca40}). To these figures one can apply the same considerations
made for the carbon data, although the preference for the weaker ph effective
potential is even more clear now, since in heavier nuclei collective effects
tend to be stronger. Indeed, the sensitivity of the response functions to
variations in the input potential is a little more pronounced in $^{40}$Ca than
in $^{12}$C, being about $25\div35\%$ for a $50\%$ change of the potential in
the $S=T=0$ channel, to be compared to the $20\div30\%$ sensitivity found in
$^{12}$C (see the discussion of Fig.~\ref{fig:RPA}).
Also the two-step contribution is, as expected, larger in $^{40}$Ca than in
$^{12}$C.

\section{ Conclusions }
\label{sec:concl}

From the discussion in the previous Section, it appears that the model presented
here gives a good description of the quasielastic $K^+$ data, the main
uncertainty being related to the strength of the effective ph interaction in the
scalar-isoscalar channel.
A good description of the data was achieved by using the interaction of 
Ref.~\cite{Nak84} quenched by $50\%$. It was shown that the use of this
interaction with the full strength led to results in clear contradiction with
the data. Indeed, there are theoretical indications that the strength of this
interaction in the scalar-isoscalar channel should be reduced.
On the other hand, from Figs.~\ref{fig:C12} and \ref{fig:Ca40} one can see that 
the $K^+$ data can be used to constrain the strength of the ph potential for 
the $S=T=0$ mode.

There are, however, two issues that need to be commented upon, namely the
reported high experimental values for the effective number of nucleons 
$N_{\text{eff}}$ \cite{Kor93,Kor95a} (see the Introduction) and the comparison
with the available calculations for this reaction that employ relativistic
dynamical models \cite{Kor93,Kor95a,Pie95}.

Concerning the first point, it might appear curious that the ``experimental''
value for $N_{\text{eff}}$ quoted in Refs.~\cite{Kor93,Kor95a} be $\sim30\%$
higher than the one calculated in the Glauber model, since we have seen that the
latter gives a nice estimate of the quasielastic cross sections. A source of
error may of course be given by the use of the effective number approximation
(\ref{eq:Neff-approx}); however, we believe that the main reason for the
overestimate of $N_{\text{eff}}$ lies in the way followed to extract the
effective number from the data \cite{Kor95a}. Indeed, $N_{\text{eff}}$ has been
obtained  by integrating the quasielastic data over the whole range of 
transferred energies: In order to do this, the data have been fitted with 
Gaussian distributions and, although a subtraction of the elastic scattering and
low energy excited levels contributions has been attempted, {\em no subtraction 
for any high energy background has been included}.
From Figs.~\ref{fig:C12} and \ref{fig:Ca40} it appears that the very high energy
tail of the cross sections is underestimated by the calculations, both for the
free and the correlated models (remember, from the discussion of 
Sec.~\ref{sec:results}, that while RPA correlations sizably affect the low
energy side of the QEP, they have little influence on the high energy part).
The data actually seem to show some structure beyond the ph response region: In
the case of quasifree electron scattering, the strength in this region is
attributed to effects beyond RPA (such as meson-exchange currents or 2p--2h
excitations).
Note also that a simple, ``straight line'', estimate of the background brings 
the value of $N_{\text{eff}}$ in close agreement with the Glauber estimate
\cite{Kor95b}.

Meson-exchange currents effects tied to the interaction of the kaon with the
nuclear pion cloud have been evaluated in Refs.~\cite{Jia92,Gar95} and found to
lead to small corrections to the $K^+$-nucleus total cross section.
The consideration of contact terms using chiral Lagrangians makes the
corrections even smaller \cite{Mei95}, so that this source of corrections cannot
account for the strength at large values of $\omega$. On the other hand, the
tail of the $\Delta$ excitation reaches this region, since the $\Delta$ acquires
a finite width, even below the pion production threshold, due to the 
$\Delta N\to NN$ reaction in the nucleus. Hence, the channel of kaon induced
$\Delta$ excitation in nuclei should become a target of both theoretical and
experimental investigation in order to further clarify these issues.

Concerning the calculations with relativistic models of 
Refs.~\cite{Kor93,Kor95a,Pie95}, it is obviously difficult to make a comparison
with our results, since in those papers the RPA response functions (calculated
in a variety of models) have been multiplied by the ``experimental''
$N_{\text{eff}}$, so that their strength is just fitted to the data (of course,
the use of the calculated value for $N_{\text{eff}}$ would result in a general
underestimate of the cross sections).
However, it should be noticed that the relativistic RPA isoscalar response is 
{\em quenched}, in contrast to our non-relativistic RPA response, which is
mainly shifted to the left and enhanced at very low energies.
Then, it is clear that this quenching has to be compensated by a higher
$N_{\text{eff}}$.
Also to be noticed is the fact that the relativistic RPA calculations are able
to describe (albeit through a fit of the total cross section) only the data at
high momentum transfers. At $q=290$ MeV/c, the large enhancement in the cross
section visible at moderate energy transfers ($15\lesssim\omega\lesssim40$) is
clearly not predicted, even in a finite nucleus calculation \cite{Kor93}, in
contrast to our non-relativistic results.

Of course, before drawing more firm conclusions one should test the model also
against the ($e,e'$) data and calculations in this direction are in progress.
In this connection, it is interesting to note that a recent reanalysis of the
($e,e'$) world data \cite{Jou96} seems to rule out the long-standing problem of
the missing strength in the charge channel, which had been interpreted in terms
of a quenching of the charge response.

\acknowledgements

We would like to acknowledge R. J. Peterson for sending us detailed information
on the experiments. This work has been supported by the program of Human Capital
and Mobility of the EU, contract n. CHRX--CT 93--0323, and by CICYT, contract
n. AEN 96--1719.

\end{document}